\documentclass[5p]{elsarticle}
\usepackage{natbib}

\usepackage{amsmath} 
\usepackage{amssymb,latexsym} 
\usepackage{commath} 
\usepackage{aas_macros}
\usepackage{threeparttable}
\usepackage{algorithm}
\usepackage{algorithmic}
\usepackage{listings}
\usepackage{graphicx}

\usepackage{hyperref}

\usepackage{xcolor}

\renewcommand{\vec}[1]{\boldsymbol{#1}}

\newcommand{\uvec}[1]{\hat{\vec{#1}}} 
\newcommand{\mat}[1]{\boldsymbol{\mathbf{#1}}}
\newcommand{\RNum}[1]{\uppercase\expandafter{\romannumeral #1\relax}}

\begin{document}


\title{Data Processing Pipeline For Tianlai Experiment}
\author[1,2,3]{Shifan Zuo}
\ead{sfzuo@bao.ac.cn}
\author[1,2]{Jixia Li}
\author[4]{Yichao Li}
\author[5]{Das Santanu}
\author[6]{Albert Stebbins}
\author[7,8]{Kiyoshi W. Masui}
\author[9,10]{Richard Shaw}
\author[11]{Jiao Zhang}
\author[1]{Fengquan Wu}
\author[1,2,12]{Xuelei Chen\corref{cor1}}
\ead{xuelei@cosmology.bao.ac.cn}
\address[1]{Key Laboratory for Computational Astrophysics, National Astronomical Observatories, Chinese Academy of Sciences, Beijing 100101, China}
\address[2]{School of Astronomy and Space Science, University of Chinese Academy of Sciences, Beijing 100049, China}
\address[3]{Department of Astronomy and Center for Astrophysics, Tsinghua University, Beijing 100084,China}
\address[4]{Department of Physics and Astronomy, University of the Western Cape, Robert Sobukwe Road, Belville 7535, Republic of South Africa}
\address[5]{Department of Physics, University of Wisconsin Madison, 1150 University Ave, Madison WI 53703, USA}
\address[6]{Fermi National Accelerator Laboratory, P.O. Box 500, Batavia IL 60510-5011, USA}
\address[7]{MIT Kavli Institute for Astrophysics and Space Research, Cambridge, 
MA 02139, USA}
\address[8]{Department of Physics, Massachusetts Institute of Technology, Cambridge, MA 02139, USA}
\address[9]{Department of Physics and Astronomy, University of British Columbia, Vancouver, BC, Canada V6T 1Z1}
\address[10]{National Research Council Canada, Herzberg Research Centre for Astronomy and Astrophysics, Dominion Radio Astrophysical Observatory, P.O. Box 248, Penticton, British Columbia, Canada V2A 6J9}
\address[11]{College of Physics and Electronic Engineering, Shanxi University, Taiyuan, Shanxi 030006, China}
\address[12]{Center of High Energy Physics, Peking University, Beijing 100871, China}

\cortext[cor1]{Corresponding author}

\begin{abstract}
The Tianlai project is a 21cm intensity mapping experiment aimed at detecting dark energy by measuring the baryon acoustic oscillation (BAO) features in the large scale structure power spectrum.  This experiment provides an opportunity to test the data processing methods for cosmological 21cm signal extraction, which is still a great challenge in current radio astronomy research.  The 21cm signal is much weaker than the foregrounds and easily affected by the imperfections in the instrumental responses. Furthermore, processing the large volumes of interferometer data poses a practical challenge. We have developed a data processing pipeline software called {\tt tlpipe} to process the drift scan survey data from the Tianlai experiment.  It performs offline data processing tasks such as radio frequency interference (RFI) flagging, array calibration, binning, and  map-making, etc. It also includes utility functions needed for the data analysis, such as data selection, transformation, visualization and others. A number of new algorithms are implemented, for example the eigenvector decomposition method for array calibration and the Tikhonov regularization for $m$-mode analysis. In this paper we describe the design and implementation of the {\tt tlpipe} and illustrate its functions with some analysis of real data.  Finally, we outline directions for 
future development of this publicly code. 
\end{abstract}

\maketitle

\section{Introduction}\label{S:intro}

The Tianlai (Chinese for ``heavenly sound'') project \citep{Chen2012,Xu:2014bya} is an experimental effort to make intensity mapping \citep{2008PhRvL.100i1303C} observations of the redshifted 21~cm line of neutral hydrogen in order to make a high fidelity 3D map of the northern sky.  From the baryon acoustic oscillation (BAO) features in the large scale structure one can determine the expansion history of the universe and probe the nature of dark energy.  The project plan include  three stages - Pathfinder, Pathfinder+, and Full Array. Presently the project is in the Pathfinder stage.  This includes a 16 element dish array, compactly arranged in two concentric rings \citep{wu2020tianlai,Zhang:2016whm}; and a cylinder array with three north-south oriented cylinders \citep{2020SCPMA..6329862L,Zhang:2016miz}, containing 31, 32, and 33 feed elements respectively. The first trial observation was made in September 2016. 

Compared with other wavebands, processing of radio data, especially that involving interferometer arrays, is often more complicated. 
For the  data-analysis process, it is often necessary to develop a customized data processing pipelines to suit some specific needs.  Writing new customized functions, when the data structures are not suitable, may make the things even more complicated. There are a number of frequently used  data processing packages for radio interferometer arrays, e.g.  {\tt Miriad} \citep{1995ASPC...77..433S} and {\tt CASA} \citep{2008ASPC..394..623J}.  These are large packages with their self-defined data formats. However, these software packages are not suitable for all arrays or for all types of observations. We do not require all their functionalities, and the overhead of learning and using these packages is quite high. Developing new data processing tasks based on them require deep knowledge of the internal coding of those packages. As Tianlai is a long term project. Presently its in the Pathfinder stage, which will later be extended to the full array. Therefore, it is always important to have a custom-made software package, which can be modified according to the system requirement. This will allow Tianlai data users to directly download the latest version of the package and apply some standard processing on the data set without much knowledge of the internal coding or the present state of the arrays.  
We note that a number of new softwares for radio astronomical data processing are also being developed presently, e.g. {\tt codex-africanus} \citep{arras2020efficient},  {\tt losoto} \citep{de2019systematic}, {\tt caracal}\citep{jozsa2020caracal}.

The Tianlai array has an unusual antenna form (cylinders) and operates in a special observing mode (drift scan). General purpose radio data processing software, such as {\tt CASA}, are quite cumbersome as they are designed to perform a large variety of tasks for a wide range of antenna configurations and observation strategies. 
None are well suited to our array without extensive customization. As such, we have developed a customized data processing pipeline software for Tianlai, named {\tt tlpipe}, which does incorporate many standard algorithms. We have coded many tasks into the pipeline to satisfy the data processing needs of the Tianlai array: reading the raw observing data, RFI flagging, calibration, data binning, final map-making, and also data selection, transformation, visualization and several other utilities.
 
 The Tianlai array raw data files are written in the HDF5 format\footnote{\url{https://www.hdfgroup.org}} \cite{Fortner1998}, and the pipeline will also output the processed data (e.g. the calibrated visibilities) or data from any intermediate step all into HDF5 files. Improper choice of file format may make the handling of large amount of data cumbersome and slow. Compared with some data formats developed at earlier times (e.g. the FITS format), the HDF5 format has much more flexibility, supporting data chunking, external (i.e. distributed) object storage, and a filter pipeline for data compression.  The HDF5 format is particularly well suited for rapidly processing large data sets as it supports parallel I/O operations.  
Also we note that the underlying data model of FITS can be easily ported to HDF5 \cite{Price2015}. A sample HDF5 data file header which provides meta information about the data is shown in Table~\ref{tab:header} in the Appendix.

The pipeline is written in the Python Programming language. The {\tt tlpipe} was initially built in 2015, when Python 3 was not yet stable, so it was developed in  Pyton~2.7. Currently we are working on updating it to Python 3. It is developed by using the MPI-framework\footnote{\url{https://www.mpi-forum.org}} \cite{Gropp1999}, so that it can be run in parallel.  However, for testing on a small data set it can also run in non-MPI environments.  Each MPI process computes part of the data set, usually a slice of the visibilities naturally partitioned along a specific axis (time, frequency, baseline) or a combination of several such axes. We have run it with as many as $10^{4}$ MPI processors without experiencing any technical difficulty.  Some performance-critical parts are statically compiled using the Cython \cite{Behnel2011} package to improve the efficiency.  The code package is developed using the git version control tool.
We have followed the PEP 8\footnote{\url{https://www.python.org/dev/peps/pep-0008/}} code style and the PEP 257\footnote{\url{https://www.python.org/dev/peps/pep-0257/}} docstring conventions, specifically we have used the {\tt numpydoc} docstring style\footnote{\url{https://numpydoc.readthedocs.io/en/latest/format.html}}, so a documentation of the package can be automatically generated by using the {\tt Sphinx}\footnote{\url{https://www.sphinx-doc.org/en/master/}} tool. Most of the code are covered by the unit test with Python {\tt unittest}\footnote{\url{https://docs.python.org/2/library/unittest.html}} framework, and Travis CI\footnote{\url{https://travis-ci.org}} is used for continuous integration.  It is an open-source project with the GNU General Public License version 3\footnote{\url{https://www.gnu.org/licenses/gpl-3.0.en.html}} and can be downloaded from GitHub\footnote{\url{https://github.com/TianlaiProject/tlpipe}\label{fn:link}}. 

In this paper, we describe the data processing pipeline code package {\tt tlpipe}, and illustrate its functions with some real data.  The package works for both the dish array and the cylinder array of the Tianlai experiment. However, in this paper we use data from the cylinder array to illustrate the functionality of the code. Note that a data processing pipeline usually refers to a running program that automatically performs a sequence of actions on the data, but in this paper we shall discuss the source code package {\tt tlpipe}, which provides the basic infrastructure and tools for building and executing such running data processing program.

The paper is organized as follows: We will first give an overview of \texttt{tlpipe} --- the data processing pipeline code for Tianlai Experiment in Section~\ref{S:ov}, and then describe the data processing tasks that are implemented in \texttt{tlpipe} in Section~\ref{S:dp}. A summary of the paper is given in Section~\ref{S:con}. A few example input parameter files are given in \ref{S:ap} to show how a data processing pipeline can be created and executed with the software package.

\section{Overview of the pipeline}\label{S:ov}

\begin{figure}[htbp]
\centering
  \includegraphics[width=0.3\textwidth]{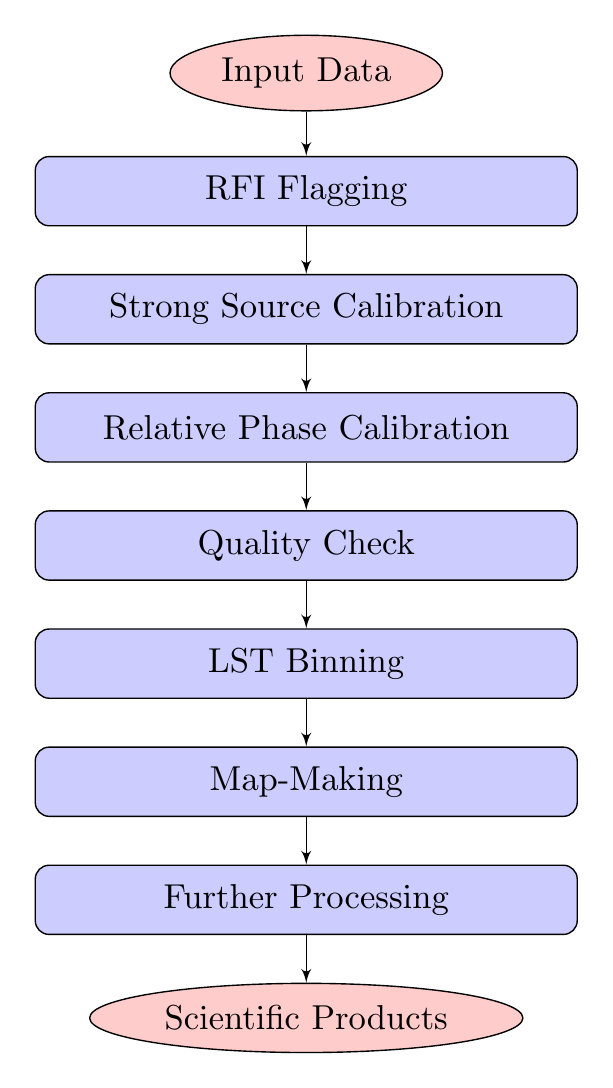}
  \caption{A schematic of the data processing procedure. } 
  \label{fig:flow} 
\end{figure}

The digital correlator produces short time integrated cross-correlations of array element receiver voltages (visibilities) which are stored on hard drive arrays in HDF5 files for off-line processing.  The Tianlai data processing pipeline code package, {\tt tlpipe}, is designed to process the data to produce 3D sky maps.  The general procedure of Tianlai data processing is shown in Fig.~\ref{fig:flow}. It has the following functions:
\begin{itemize}
\item{Input Data.} The radio interferometer array outputs large amount of data. For each integration time (currently set at $\sim 4$s for the cylinder array), the digital correlator produces 
$N_{\nu}N_{rx} (N_{rx} +1)/2 $ visibilities, where $N_{rx}$ is the number of receiver elements, and $N_{\nu}$ is the number of frequency channels. The Tianlai correlator saves the visibility data for every 10 minute interval into a new HDF5 file. The size of each such file is about 21 GB. The {\tt tlpipe} can read one or a number of such files for data processing.

\item{RFI Flagging.} Radio frequency interference (RFI) is identified  by its unusual magnitude (much larger than typical astronomical source or noise) and flagged so that the data contaminated by RFI will be ignored in further processing. In practice, a {\it mask} file is generated to record the times and frequencies contaminated by RFI. 

\item{Relative Phase Calibration.} There are only a few strong sources in the sky which can be used for calibration due to the relatively low sensitivity of the individual array elements. In order to maintain the instrumental phase of the array elements a relative phase calibration is done every few minutes by employing an artificial noise source. The noise source is put on a nearby hill. It broadcasts broad band noise signals for one or a few integration cycles every few minutes. The magnitude of the calibrator signal is adjusted to be 
higher than the astronomical sources, but still within the linear range of the receivers. The {\tt tlpipe} extracts this part of the data for calibration computation, and afterwards also flags them.

\item{Quality Checks.}  At this point, a number of data quality checks can be performed by examining the variation of the complex gains for the different receivers. Outlying data may be flagged and analyzed more carefully before being used. 

\item{LST Binning.} After calibrating the visibilities data from different days of observations are averaged and re-binned according to the local sidereal time (LST). For a drift scan telescope nearly all astronomical sources are periodic day by day (except the Sun, etc). Non-periodicity can be used to assess the stability of the system. Visibilities which significantly deviate from other days are removed. The re-binned data set can be used for $m$-mode analysis.

\item{Map-Making.} A 3D sky map is made using $m$-mode analysis (2D angular position $\times$ 1D frequency or 21cm redshift).

\item{Further Processing.} Scientific data products can be made by further processing; such as identifying point sources, making 21cm maps via foreground subtraction, and deriving the 21cm power spectrum.

\end{itemize}

The array generates a large amount of data in its normal observation.  Processing such large amounts of data within an acceptable time  needs high performance parallel computing.  
we have chosen the widely used MPI architecture. Specifically we have used the data parallel model, in which tasks are assigned to processes and each task performs similar types of operations on different data. As will be described in Section~\ref{S:co}, data parallel model is well suited for the data processing requirements, as many of the data processing tasks naturally works for a slice of the visibility data along one axis or several axes. The disadvantage of this design is that the data load may not be balanced for all processes, so some processes may waste time waiting for others during a synchronization. To reduce the impact on the performance, the data are divided as evenly as possible during the pipeline execution. During the development, we have taken special care to ensure the package works perfectly even in a non-MPI environment so one can run the jobs on a single processor for smaller data sets.

\subsection{Code File Organization}

The tlpipe is designed in a modular fashion. The package is organized in sub-packages or modules according to functionality.  The main executing framework, the data processing tasks, and the customized data containers are organized as separate classes with clear inheritance hierarchy
by using an object oriented programming (OOP) \cite{Rentsch1982} scheme.  OOP is a commonly used programming paradigm that relies on the concept of classes and objects. It is used to structure a software program into simple, reusable pieces of code blueprints (usually called classes) which are used to create individual instances of objects.  OOP can help to manage complexity in large programs.  An object can package data and the operations on it so that only the operations are publicly accessible and internal details of the data structures are hidden.  This information hiding makes large-scale programming easier by allowing a programmer to think about each part of the program in isolation.  In addition, objects may be derived from more general ones, “inheriting” their capabilities.  Such an object hierarchy makes it possible to define specialized objects without repeating all that is in the more general ones.  Python supports OOP well.  Many of the dependent packages that {\tt tlpipe} depends on, for example, {\tt AIPY}\footnote{\url{https://github.com/HERA-Team/aipy}} \cite{Parsons2016}, {\tt caput}\footnote{\url{https://github.com/radiocosmology/caput}} and so on are built also on an OOP paradigm. By inheriting from the classes in this packages, a lot of efforts can be saved.  Auxiliary utilities are coded as individual functions to be called by other components. The modular design is easier to understand, allows developers to focus on the component of interest, and minimizes the impact of modification of one component on other components. 

\begin{description}
  \item[core:] The definition and implementation of the class for the primary beam model of the antennas (including both dish and cylinder type) 
  and the interferometer array. They are modeled as classes, based on the classes (Beam, Antenna, AntennaArray) defined in the {\tt AIPY}\footnote{\url{https://github.com/HERA-Team/aipy}} \cite{Parsons2016} package. 
  \item[pipeline:] Pipeline control including the task executing framework, also called the task manager, and the base functionality for the tasks.
  \item[container:] The customized data containers.
  \item[timestream:] Tasks for processing the observed time stream data, such as {\tt rfi\_flagging.py} for RFI flagging,\\ 
  {\tt ps\_cal.py} for calibration by using a strong point source on the sky, {\tt map\_making.py} for map-making, and so on. 
  \item[rfi:] RFI flagging algorithms.
  \item[cal:] A catalog of calibrator sources and  things related to calibration.
  \item[map:] The $m$-mode map-making method.
  \item[foreground:] For foreground removal.
  \item[powerspectrum:] For power spectrum estimation.
  \item[plot:] Visualization utilities.
  \item[utils:] Miscellaneous functions used in the code.
  \item[test:] Unit testing code for the package.
\end{description}

The data processing  is a multi-step process, and during the actual data processing and analysis,  various tools for data exploration and transformations are needed. We abstract the data processing as a  number of individual tasks, and the pipeline is made by executing a series of such tasks using a common task executing framework.  All tasks are coded with a consistent interface, which is achieved by first defining an abstract base task class in which a few standard method interfaces are defined, then the specific tasks are inherited from the base class or its sub-class, with implementations of the methods. In this way the different tasks are called and executed by the task manager. 

Parameters needed by the various tasks are not  hard-coded, instead they are fed to the pipeline by an input parameter file. The user can specify the list of tasks to be executed, their order, and all the parameters different from their default values in the parameter file. The idea is to make the input file as the only input to the pipeline and supply the parameters through the input file.  The parameters are parsed and assigned to the corresponding tasks before the execution of the tasks. This is very  flexible, avoids manually changing the code of the package, and allows simultaneously multiple run of the package for different data processing tasks by specifying different input parameters. We have incorporated a modified version of the {\tt kiyopy}\footnote{\url{https://github.com/kiyo-masui/kiyopy}} package for parsing of the parameters.  The main modifications are those which make it work smoothly in a multi-process MPI environment. The execution framework and the data containers are based on the infrastructure developed in the {\tt
 caput}\footnote{\url{https://github.com/radiocosmology/caput}} package.
These were developed for the Green Bank Telescope (GBT) Intensity Mapping project \cite{2013ApJ...763L..20M,Switzer2013,Switzer2015} and the Canadian Hydrogen Intensity Mapping Experiment (CHIME) 
\cite{2014SPIE.9145E..22B}.

\subsection{Execution Framework}\label{S:co}

\begin{figure}[htbp]
  \includegraphics[width=0.45\textwidth]{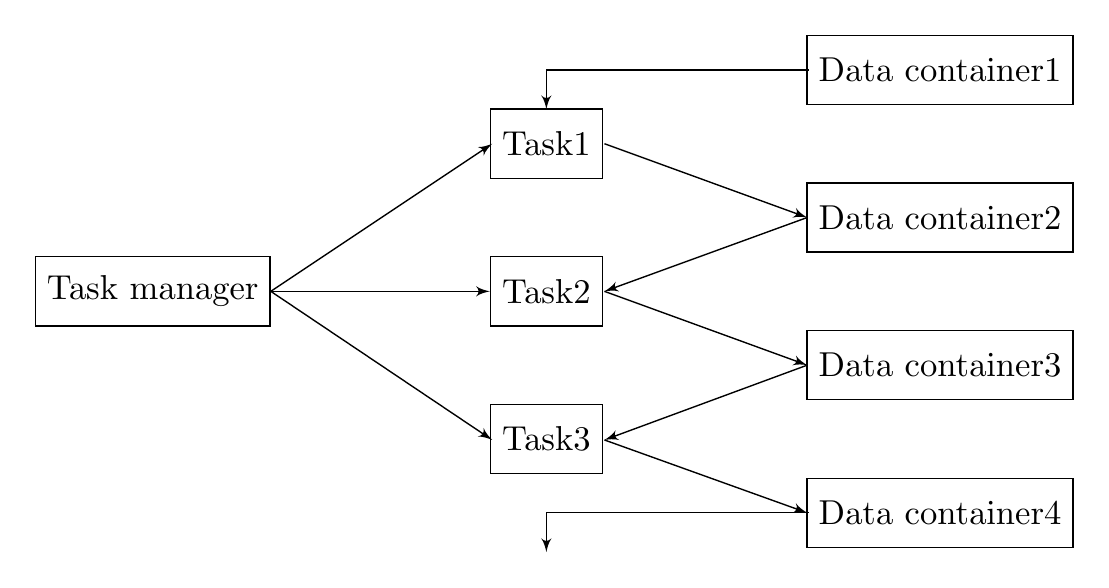}
  \caption{Schematic relation between
the three major components of {\tt tlpipe}: the task manager, the tasks, and the data
container. } 
  \label{fig:3c} 
\end{figure}

The {\tt tlpipe} package consists of three main components that interact with each other: the task execution framework, the tasks, and the data container.  The task execution framework (the task manager) controls the execution of the tasks. A task is  an individual or independent data processing step.  It is implemented according to a few common interfaces so that it can be executed by the task manager.  The data container holds the data and some descriptive metadata to be processed by a task. A data processing pipeline usually consists of several tasks  which are executed in a specified order.  The same tasks may be executed multiple times depending on data processing requirement.  The set-up and execution of tasks are controlled by the task manager according to an input parameter file provided by the user. The input parameters file fully determine the pipeline control flow and the behavior of each task in the pipeline.  Full data processing can be submitted and run on a personal computer, a cluster, or a supercomputer with a single parameter file. No other human intervention is required until it is finished or stops due to some error.
The three major components are shown in Fig.~\ref{fig:3c}. 

The task manager first reads in the input parameter file and parses it.  The tasks listed in the file are put into a list according to their execution order.  The task manager initializes an instance for each task by using the parameters specified in the parameter file and then executes all the tasks  according to control flow. Once a task is completed, the task manger takes output and passes it to the next task as an input.  Once all the tasks are done, the task manager deletes the instances and finishes the pipeline.  See Algorithm~\ref{alg:exe} for the algorithm that {\tt tlpipe} follows to execute tasks.

\begin{algorithm}
\caption{The algorithm that {\tt tlpipe} follows to execute tasks.\label{alg:exe}}
\begin{algorithmic}[1]

  \STATE Read in the input parameter file and parses it.
  \STATE Put the tasks listed in the file into a list according to their execution order.
  \STATE Initializes an instance for each task by using the parameters specified in the parameter file.
  \FOR{ task in task list}
    \IF{no input data container}
      \IF{a list of data files is given}
        \STATE Create a data container by loading data from the specified data files as the input.
      \ELSE
        \STATE Raise an exception.
      \ENDIF
    \ELSE
      \STATE Pass the input data container to the task.
      \STATE Execute the task.
      \STATE Take the output data container and maybe pass it to the next task.
    \ENDIF
    \ENDFOR

  \STATE Deletes the instances and finishes the pipeline.

\end{algorithmic}
\end{algorithm}

A task is defined by inheriting from a common base task class, which defines the standard methods and their interfaces.  The three methods defined by  the base task class are {\tt setup()}, {\tt next()} and {\tt finish()}, which the subclasses should implement for their specific job.  
The methods {\tt setup()} and {\tt finish()} are called by the task manager only once, for setting-up and finalizing the task.  Usually we  do not need to override these in the  sub-classes if there is no special or unusual things to do.  In such case the default behavior will be executed.  The method {\tt next()} usually needs to be overridden to perform some specific job.  It may be called by the task manager multiple times for   executing some tasks. As shown in Fig.~\ref{fig:3c}, a data processing task usually accepts a data container as input and returns another data container as output.  In case there is no data container for input, if a list of data files are given, the task will first create a data container by loading data from the specified data files and use it as input, otherwise, an exception will be raised.  The input and output data container can be the same, with or without modification of the  data in it. The task can call general data operation interfaces provided by the data container to operate on the data. Most of the tasks can be found at the {\tt timestream} and
{\tt plot} directory of the code repository.
New tasks will be added as we continue developing \texttt{tlpipe}.

The visibility data, various auxiliary data and meta data are grouped together in a data container. The data container provides general data operation interfaces.  The tasks access the data by using these interfaces, so that the inner consistency of the data is maintained.  The data container is combined with the data processing tasks, usually as an input to the task, and as an output after the operations are finished by the task.

Two data containers are used in the {\tt tlpipe} package: the {\tt RawTimestream} data container, that holds the the raw visibility data with mixed polarization and baseline, and the {\tt Timestream} data container, that holds the visibility data with polarization and baseline separated.  Both data containers hold the same data, but in two different formats. Some of the tasks can be processed easily in a particular type of data container.  One can convert a {\tt RawTimestream} data container into a {\tt Timestream} data container.  Most tasks can take either of the data containers as input, but some tasks can accept only one of the two.   The data container provides a general data operation interface for the tasks to act on the data.  It allows a task to operate on the data, either as a whole by calling 
{\tt all\_data\_operate()} 
method, or on a specific axis of the visibilities by calling methods like
{\tt time\_data\_operate()}, 
{\tt freq\_data\_operate()}, 
{\tt
bl\_data\_operate()}, 
etc.  We can also operate a task on a combination of several axes of the visibilities by calling its methods, such as 
\\{\tt time\_and\_freq\_data\_operate()}, etc.

Under the MPI environment, methods that allow data to be operated on a specific axis or a combination of several axes, automatically split the data set along the corresponding axis or the combination of axes into data slices.  Each slice is then sent to an MPI process  for processing.   For example, if there are $N_{p}$ MPI processes, by calling the {\tt time\_data\_operate()} method to act on a visibility dataset with $N_{t}$ time points, the method internally splits the dataset into $N_{p}$ parts along the time axis. Each part gets
$N = \left \lfloor{\frac{N_{t}}{N_{p}}}\right \rfloor $ 
time points. Some parts may have $N + 1$ time points if $N_{t}$ is not a multiple of $N_{p}$.
Tasks choose the most natural data splitting and parallelism schedules according to their job  by calling appropriate data operation methods of the data container. For example, the SumThreshold RFI flagging task is naturally done baseline by baseline on a time-frequency visibility slice. So it calls the {\tt bl\_data\_operate()} method.

\section{Data Processing Tasks}\label{S:dp}

In {\tt tlpipe} we have implemented the basic data processing procedures, such as data reading, RFI flagging, noise source calibration, point source calibration, data binning, and map-making etc., as well as  various auxiliary analysis utilities like data selection, transformation, visualization and so on.  Here, we briefly describe the main  processing tasks implemented in {\tt tlpipe}. 

\subsection{Data Reading and Pre-Processing}

The observational data are naturally divided in the time dimension and saved as a separate HDF5 file.
The observed visibilities are saved as a three-dimensional data sets (i.e., array) with its first dimension time, second dimension frequency and last dimension the {\it digital channel} (the cross-correlation or auto-correlation for a pair of polarized feed signal from a baseline generates the signal for one digital channel). Each file contains about 10 minutes observation data for the cylinder array with a size of about 21 GBs.  Other meta data is also saved as data sets or attributes in the HDF5 file.  Each file contains complete information of the observation during this time period. Therefore, each data file can be separately processed without reference to others.

While analyzing the data using the pipeline, one selects particular time points, frequency bins or a list of feeds by setting the corresponding parameters, such as {\tt time\_select}, {\tt frequency\_select}, {\tt feed\_select}, in the parameter file.  The analysis pipeline reads only the selected chunks of data from the given data files. If these parameters are not explicitly set, the code, by default, reads all data from the given data files. Thus, if one wants to make a trial analysis on a small amount of data contained in a few files, either to check a new analysis procedure or to see the quality of the data, there is no need to explicitly set these parameters every time.  For drift scan observation, the natural data processing time unit is a sidereal day.  Therefore, if the input data is longer than a sidereal day, the pipeline will automatically split the data in length of a sidereal day and process each data chunk.  The user can freely select how much data he/she wants to process each time by changing the value of the corresponding parameter, in the input parameter file. 
  
The selected data is loaded into a data container, usually the {\tt RawTimestream} data container, for analysis. It can then be converted to be a {\tt Timestream} data container according to one's data processing needs.  The data container mirrors all the corresponding HDF5 file objects (specifically Python HDF5 package {\tt h5py} objects), such as the group, the data set, and the attribute, and preserves their corresponding organization and structural relations in memory.  This correspondence enables the user to apply familiar HDF5 file operation methods on the data container, and also makes the data I/O conversion between the data container in memory and the corresponding HDF5 files on disk simple. 

We also generate a Boolean Mask array, having the same dimensions as that of the visibility array, in the data-container. If there is any missing data due to broken hardware or any other errors, we set the mask value corresponding to those visibilities to True.

\subsection{RFI Flagging}\label{S:rfi}

Several RFI flagging methods are implemented in the
{\tt tlpipe} package.  However, our tests show that a two step RFI flagging process works well for Tianlai data. 

In the first step, we use the SumThreshold method
\citep{Offringa2010}, which has been originally used by the LOFAR team.  The SumThreshold method works on the background subtracted residual visibility data. We first fit a smooth background surface for the amplitude of the time-frequency visibility for each baseline and get the residual by subtracting it from the visibility.

As discussed earlier, in Tianlai we use an artificial noise source in order to track the variation of instrumental phase.  The noise source broadcasts strong noise signals in regularly spaced intervals. These regular calibrator signals in the data make it hard to fit a smooth background.   We solve this problem by identifying the parts of data, which contain such calibrator signals, and then replacing those by cubic spline interpolation. A smooth background is fitted after that.  Several background fitting methods have been implemented, e.g. moving local average, moving local median,  low order polynomial fitting, Gaussian high-pass filter, etc.  In practice, we found that by setting appropriate parameters, the Gaussian high-pass filter works best. So we mostly use this filter during data analysis.  

The SumThreshold method uses the sum of a combination of one or more consecutive data points along a time or a frequency axis as a threshold criterion. The data samples are flagged as RFI if their sum exceeds the threshold 
$\chi_{i} =
  \frac{\chi_{1}}{\rho^{\log_{2} i}}$, 
where the first threshold $\chi_{1}$ is determined by the Median Absolute Deviation (MAD) of the data,  $\rho$ is empirically set as 1.5, and $i$ is the number of the data samples considered. The MAD for a data set $X = \{X_1, X_2, \dots, X_n\}$ is defined as $\text{median}(|X - \text{median}(X)|)$. It is a robust measure of the variability of the data samples in presence of some outlier corruption.
  
  \begin{figure}[htb]
\centering
  \includegraphics[width=0.45\textwidth]{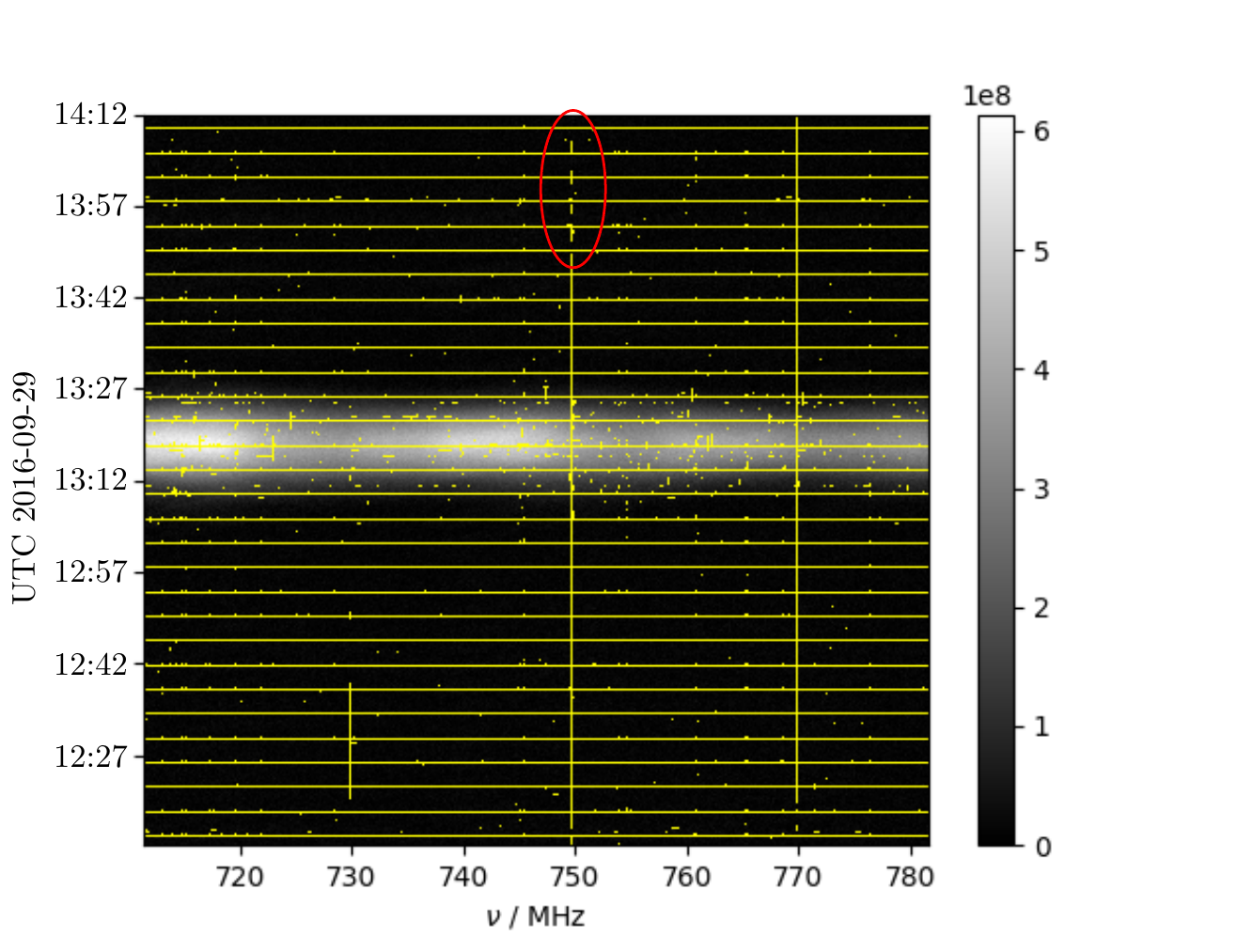}
  \includegraphics[width=0.45\textwidth]{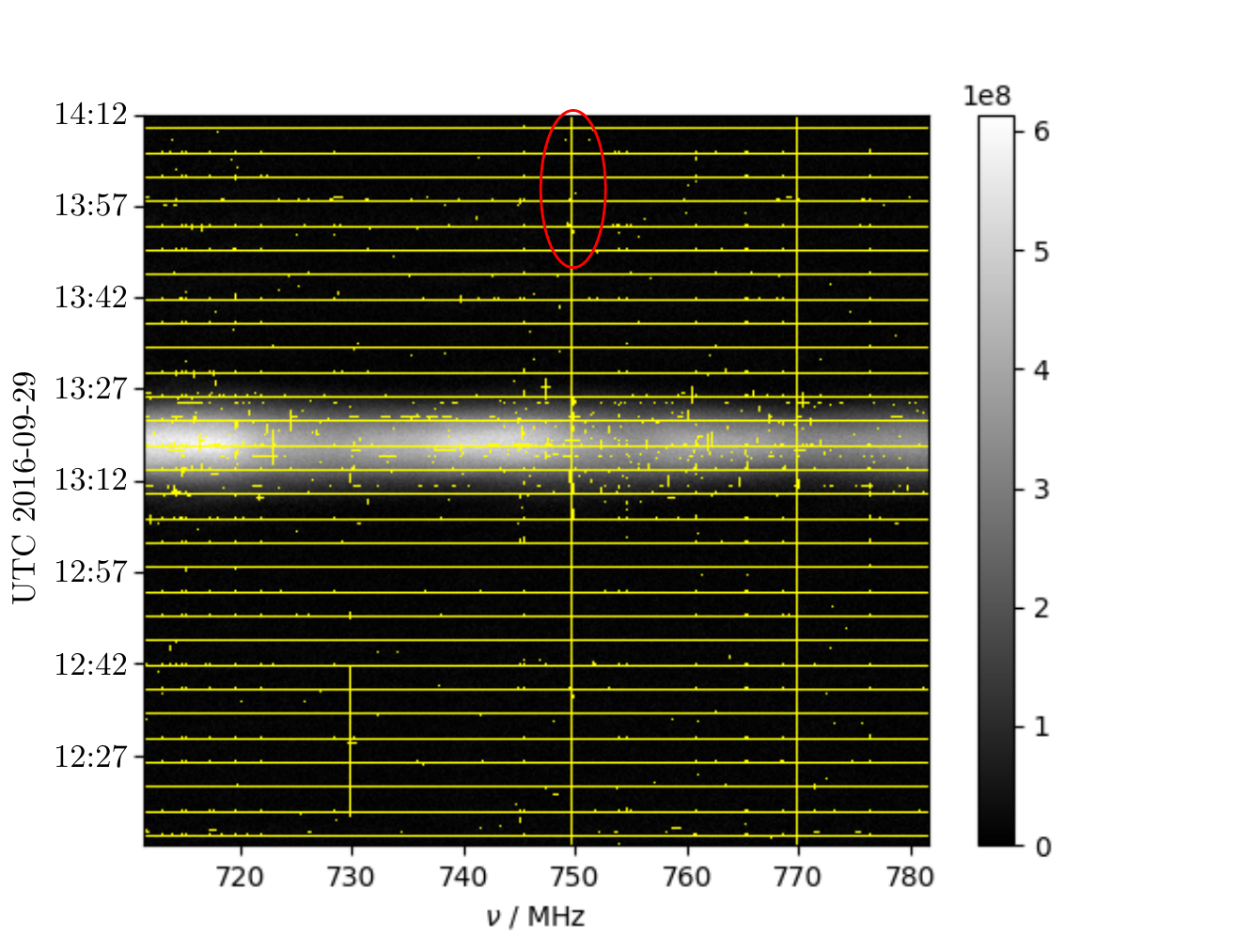}
  \caption{RFI flagging result by using the SumThreshold method
    (top) and the SIR operator (bottom). The underlying image of each sub-figure shows the amplitude of the visibilites observed by one baseline of the Tianlai cylinder array in arbitrary unit (un-calibrated visibilites). The bright belt in the center of the image is generated by the strong signal of Cygnus A. The image is covered by yellow colors which shows the position of the flagged values. The equally spaced horizontal yellow lines are positions of the much more stronger signal generated by a periodically broadcasted noise source used for calibration.}
  \label{fig:rfi} 
\end{figure}

To further improve the detection of RFI, we then apply the scale-invariant rank (SIR) operator method \citep{Offringa2012}.  This method uses a one-dimensional mathematical morphology technique to find adjacent intervals in the time or frequency domain that are likely to be affected by RFI. Specifically the scale-invariant rank operator will flag a subsequence when more than $(1 - \eta) N$ of its samples are flagged, with $N$ the number of samples in the subsequence and $\eta$ a constant, $0 \le \eta \le 1$.  The most direct effect of this
method is to extend a line segment along its two ends by a factor about $\eta$, thus fill or shrink some gaps left by the previous RFI flagging results. 

One example of the effect of applying this method is shown in the bottom panel of Fig.~\ref{fig:rfi}.  The regularly spaced horizontal yellow lines are the signals from the noise source used for the array's relative phase calibration.  Other lines and dots in yellow color are flagged data.  The white belt in the middle is the signal of the strong source Cyg A.  The SumThreshold method may miss flagging some of the weak RFI, while the SIR operator method can fill in some of the gaps between the line segments, for example shown in the red circled area.

Direct implementation of the SumThreshold method and the SIR  operator in Python is quite slow, so we have implemented the algorithm in {\tt Cython} \cite{Behnel2011}, which is then compiled as extension modules for Python. This speeds up the RFI flagging method by about a factor of 1000 times. 

In the data container the RFI flags are recorded by setting the corresponding elements of the mask array to {\tt True}.  The mask array goes along with the visibility data array in all subsequent data processing steps and allows subsequent data processing tasks to appropriately take into account the flagged values. 

\subsection{Calibration}\label{S:cal}

We have implemented a two-step calibration procedure in {\tt tlpipe}.  First, a strong radio source is used to make an absolute gain calibration.  This gives the actual amplitudes and phases of the complex gains at the time of the source transit.  Then a relative phase calibration is done by using the periodic signal from the artificial noise source signal to remove the phase variations over time. 

\begin{figure}[thb]
\centering
  \includegraphics[width=0.45\textwidth]{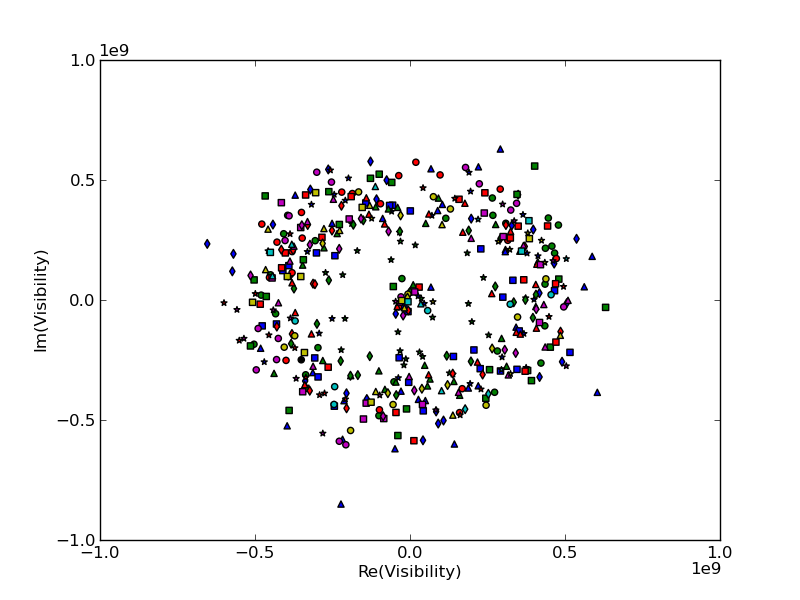}\\
  \includegraphics[width=0.45\textwidth]{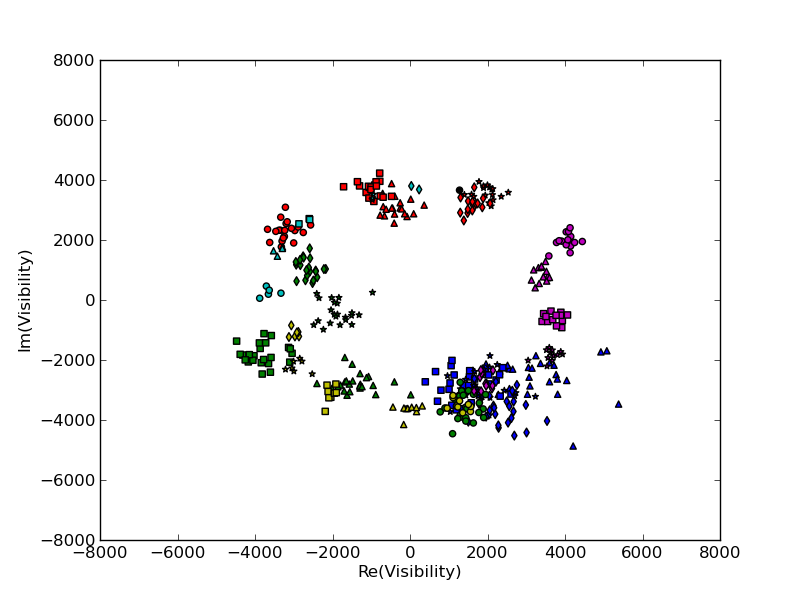}
  \caption{The redundant baseline visibilities for one of the cylinder before (top) and after (bottom) the first step calibration.  Data points with same mark and color are visibilities observed by redundant baselines (baselines with same length and direction) and should be the same in theory when there is no noise. These data points are spread out before the calibration but collapsed to clusters after the calibration.}
  \label{fig:cal} 
\end{figure}

In the first step, an eigenvector-based method is applied by using the signal of a strong point source like Cygnus A for absolute calibration. To improve the calibration precision, a stable principle component analysis (SPCA) method is used to help separate outliers and noise from the calibrator signal before the eigen-analysis of the visibility matrix \citep{Zuo2018}. The eigenvector of the outliers and noise separated visibility matrix gives the quantity
\begin{equation}
  G_i = g_i A_i(\uvec{n}_0) e^{-2 \pi i \uvec{n}_{0} \cdot \vec{u}_{i}},
\end{equation}
where $g_{i}$ is the complex gain of the receiver $i$, $A_i(\uvec{n}_0)$ is the primary beam response of this receiver in the direction of the calibrator $\uvec{n}_{0}$, and $\vec{u}_{i}$ is the position vector of this receiver. We can choose a normalization $A_{i}(\uvec{z}) = 1$ at the direction of the zenith $\uvec{z}$ to solve for the gain $g_{i}$.  The amplitude $|G_{i}|$ is proportional to the beam profile $A_i(\uvec{n}_0)$ along the transit track of the calibrator,
\[
  | G_{i} | \propto A_i(\uvec{n}_0).
\]
From this we have fitted the east-west direction beam profile of the Tianlai cylinder receiver.

In Fig.~\ref{fig:cal} we show the visibilities observed by the baselines located on one cylinder before and after the first calibration step. We see the visibilities of redundant baselines are aggregated into clusters after the calibration. Because the visibilities of redundant baselines are the same in theory, this demonstrates the effectiveness of our calibration.

We then use the periodically broadcasted artificial noise source signal to do the relative phase variation correction, so as to compensate the phase variations over time induced by the cable delay. This is done by using the difference between the signal $V_{ij}^{\text{on}}$ when the noise source is on and the adjacent signal $V_{ij}^{\text{off}}$ when the noise source is off, from which the relative phase due to cable delay $\phi_{ij}$ is obtained. Specifically we have
\begin{equation}
  \phi_{ij}=  \text{Arg}(V_{ij}^{\text{on}} - V_{ij}^{\text{off}}) = k \Delta L_{ij} + \mathrm{const.},
\end{equation}
where $k$ is the wave number, and $\Delta L_{ij}$ is the equivalent instrument delay difference between the channels $i$ and $j$.  Then we compensate for the relative phase change due to $\Delta L$ in the relative phase calibrated visibility by
\begin{equation}
 V_{ij}^{\text{rel-cal}} = e^{-i \phi_{ij}}  V_{ij}.
\end{equation}
This gives the complex gains of the receivers and the calibrated visibility data.  The gains are saved to disk for future analysis.  One can make a number of checks on the consistency and stability of the system using this processed data and estimate the accuracy of the observation. 

In the future the calibration step may be further improved by using a calibration source catalog or sky model and implementing the full-polarized calibration to correct for direction-dependent effects, or by implementing some newly developed calibration methods, etc.

\subsection{Map-Making}\label{S:map}

After calibrating the data, we can combine 
the data, taken at different times of the day or even different days, to make a sky map. 
The measured four polarization visibilities $V_{XX}$, $V_{YY}$, $V_{XY}$  and $V_{YX}$ ($X$ and $Y$ denotes the two orthogonal linear polarization of the feed) are 
combined to form 
four Stokes visibilities $V_{I}$, $V_{Q}$, $V_{U}$ and  $V_{V}$.
\begin{eqnarray}
      V_{I} &=& \frac{1}{2} (V_{XX} + V_{YY}),\\
      V_{Q} &=& \frac{1}{2} (V_{XX} - V_{YY}),\\
      V_{U} &=& \frac{1}{2} (V_{XY} + V_{YX}),\\
      V_{V} &=& -\frac{i}{2}  (V_{XY} - V_{YX}).
\end{eqnarray}
For continuous drift scan observations (array stationary), 
the signals observed at the same local sidereal time can be averaged to improve the signal-to-noise ratio. In addition, the high time resolution (short integration time) of the observation means data points observed adjacent in time are not independent and can be binned together.  We call the process, Local Sidereal time (LST) binning. 

As Tianlai array is drift scanning
we use the $m$-mode analysis for map-making~\citep{Shaw2014,Shaw2015,Zhang:2016whm,Zhang:2016miz}. 
This method makes use of the periodicity in
visibility due to the rotation of the earth to decompose the visibility into $m$-modes.  The noise in each mode should be statistically independent.  We can make maps of each of the Stokes parameter. Here for simplicity we just consider the intensity, the Stokes parameter $I$.  For each $m$-mode, we have
\begin{equation} \label{eq:vm}
  V_{(ij)}^{m} = \sum_{l} B_{(ij) \, l}^{m} a_{l}^{m} + n_{(ij)}^{m},
\end{equation}
where, $B_{(ij)l}^{m}$ and $a_{l}^{m}$ are the spherical harmonic coefficients of the beam transfer function,
$$B_{ij}(\uvec{n}) = A_{i}(\uvec{n}) A^{*}_{j}(\uvec{n}) e^{2\pi i \uvec{n} \cdot
  \vec{u}_{ij}},$$ 
and the brightness temperature of the sky, $T(\uvec{n})$, respectively.  Here $i$, $j$ are indices of a baseline $\vec{u}_{ij}$ formed by the two feeds $i$ and $j$, $\uvec{n}$ is a direction on the sky, and $A_{i}(\uvec{n})$ is the primary beam of feed $i$.  Rewriting Eq.~\ref{eq:vm} in vector-matrix form yields
\begin{equation} \label{eq:vv}
  \vec{v} = \mat{B}\,\vec{a} + \vec{n}.
\end{equation}
In general $\mat{B}$ is not a square matrix and different methods can be used to optimally solve this linear equation for $\vec{a}$.  By symmetry $\mat{B}$ is block diagonal for different $m$ modes and one can {\it independently} solve this linear equation for each block (each $m$) .
This reduces the required computation by a factor $\sim1000$ for the Tianlai Pathfinder Array.

We incorporate the {\it driftscan}\footnote{\url{https://github.com/radiocosmology/driftscan}} package, developed 
by Richard Shaw, into {\tt tlpipe}, 
with some modifications and new developments for our particular 
requirements. To improve map-making quality, we have implemented a Tikhonov regularization-based solver in addition to the existing Moore-Penrose pseudoinverse solver in the {\it driftscan} package. The Tikhonov regularization method solves the optimization problem
\begin{equation} \label{eq:tk}
  \min_{\vec{a}} \, \norm{\vec{v} - \mat{B} \vec{a}}^{2} + \varepsilon \norm{\vec{a}}^{2}
\end{equation}
with the regularization parameter $\varepsilon > 0$. The solution of the Tiknonov regularization problem is
\begin{equation} \label{eq:xtk}
  \hat{\vec{a}} = (\mat{B}^{*} \mat{B} + \varepsilon \mat{I})^{-1} \mat{B}^{*} \vec{v},
\end{equation}
where $\mat{I}$ is the identity matrix.

Besides the visibility data, map-making also requires the beam response of the telescope.  Here for illustration we  use a simple beam model, which is the product of two 1D functions: In the East-West direction, the response is calculated by illuminating the cylinder with the dipole beam, and solving for the diffraction in the Fraunhofer limit; and in the North-South direction, the response is just the feed amplitude, taking the cylinder as a reflector. For details of the beam model and its polarized response, see \citep{Shaw2015}.  Using this beam, the spherical harmonic coefficients of the beam transfer functions $B_{(ij)l}^{m}$ can be computed, or pre-computed and saved in a file which makes the map-making process faster.  In the future a better model will be obtained by a combination of electromagnetic field simulations, calibration with a drone, and
self-calibration using the observational data.

\begin{figure}[htbp]
  \includegraphics[trim=180 10 180 10,clip,width=0.4\textwidth]{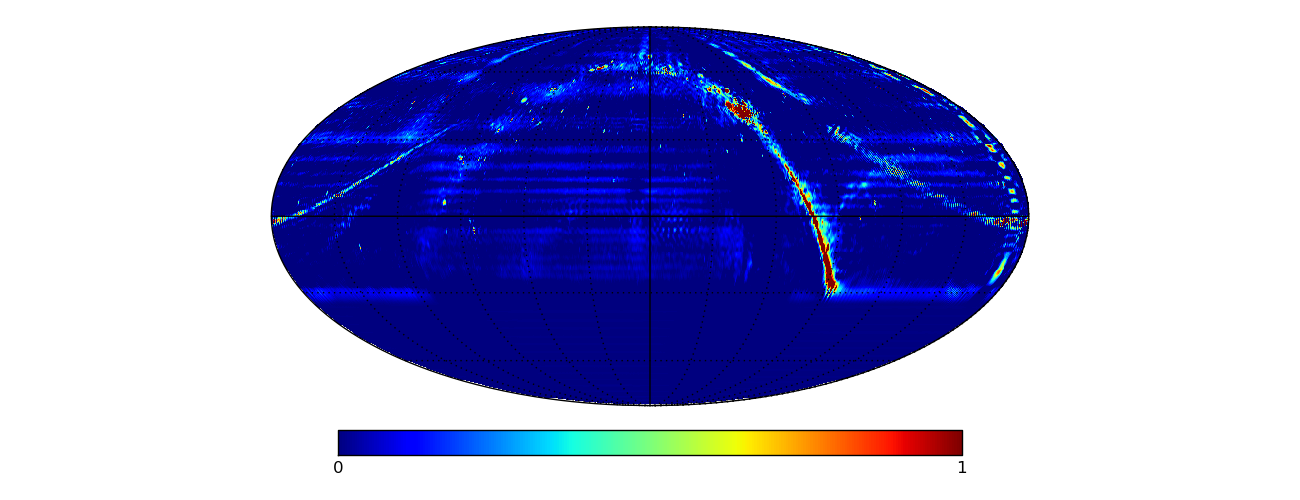}
  \caption{The resulting map (at 750 MHz) by using the Tikhonov
    Regularization based technique.}
  \label{fig:tkm}
\end{figure}

The output of the pipeline is the sky map discretized in HEALPix\footnote{\url{https://healpix.sourceforge.io}} \citep{Gorski2005} piexelization.
An example of a map, obtained from the first light observation from 2016/09/27 20:15:45 to 2016/10/02 22:36:55 Beijing time (UTC+8), is shown in Fig.~\ref{fig:tkm}.

We plan to improve the map-making code by implementing additional functions, e.g. various deconvolution algorithms. In addition to the present $m$-mode map-making function, which uses whole day observations, one may also implement instantaneous imaging function which is more useful for some transient sources. This function may be implemented with beam-forming algorithms.  

Furthermore, the sky is not entire static, there are 
radio sources whose  luminosity varies over time (e.g. pulsars and quasars), transient sources (e.g. fast radio bursts, gamma-ray burst radio afterglows),
and even sources whose position varies with time (e.g Sun and planets). At a higher level, one needs to consider how to  take into account of such sources in the map-making algorithm.  

\subsection{Other Utilities and Further Processing}

To make the package more user friendly, we provide two script commands: {\tt tlpipe} and {\tt h5info}, which can be directly used in the console after installing the package. The {\tt tlpipe} command is the one that executes the data analysis pipeline from the instructions in a parameter file provided by the user. The {\tt h5info} command inspects and outputs the summary information of one or a list of HDF5 files.  This enables the user to quickly 
scan the contents of the HDF5 data files; the HDF5 data format is used throughout the data analysis pipeline.

Further data processing can be done starting from either the calibrated visibilities or the sky maps. In principle, the two are equivalent in our data processing pipeline because the map is a lossless form of information compression of the calibrated time series visibilities under the linear map-making process (the $m$-mode method is a linear map-maker) \citep{Tegmark1997}.

We plan to implement some foreground subtraction methods in the future, for example, the PCA method  \citep{Masui2013,Sazy2015,Switzer2013,Switzer2015}, the ICA method \citep{Chapman2012,Wolz2014,Alonso2015}, the Karhunen-Lo\`{e}ve (KL) transform method \citep{Shaw2014,Shaw2015}, the robust PCA method \citep{Zuo2019}, etc.  Methods like the quadratic estimator \citep{Tegmark1997,Bond1998} will also be implemented for 21~cm signal power spectrum estimation, or to cross-correlate foreground-subtracted data with other cosmological tracers.

\section{Summary}\label{S:con}

We have presented the software package {\tt tlpipe} developed as data processing pipeline for the Tianlai array. The {\tt tlpipe} package now has a full set of function modules, which can perform basic data processing tasks such as reading data file, RFI flagging, calibration, data binning, and final map-making.  Data selection, transformation, visualization can all be done conveniently using this framework, and it is relatively straightforward to add new functions, or implement additional or optional algorithms  within the framework. It is also convenient to incorporate off-shell tool modules developed by others for specific tasks. 
The code has received a software authorship registration \citep{Zuo2017}, but is made publicly available for use with GNU license V3. The source code, a docker image and documents about the code package can be downloaded at the site given in footnote \ref{fn:link}.

\section*{Acknowledgments}

We thank Ue-Li Pen for suggestions. The development and test of the {\tt tlpipe} package was performed on the Tianhe-2 supercomputer at the  National SupercomputingCenter in Guangzhou, Sun Yat-Sen University with the support of NSFC  supercomputing Joint Grant U1501501.  We acknowledge the support of the Astronomical Technology Center of NAOC, this work is done with the support of the Tianlai cylinder array.  The Tianlai survey is supported by the MoST grant 2016YFE0100300.  This work is supported by the NSFC grant 11473044, 11633004, the NSFC-ISF joint research program No. 11761141012 and the CAS Frontier Science Key Project QYZDJ-SSW-SLH017. Work at University of Wisconsin at Madison and Fermi National Accelerator Laboratory (Fermilab) is partially supported by NSF Award AST-1616554.  Fermilab is operated by Fermi Research Alliance, LLC, under Contract No. DE-AC02-07CH11359 with the US Department of Energy.  This document was prepared by the Tianlai collaboration.

\appendix

\section{Example pipeline}\label{S:ap}

Here we show an example input parameter file for the pipeline. It
selects 30 minutes data from a list of given observational data
files (saved in HDF5 format), then finds and masks data when the
artificial noise source is on, finally makes waterfall plots (a
time-frequency plan of each baseline) of the selected data
segment. One can run the pipeline by executing the follow command
(no MPI parallelism acceleration)

{\tt \$tlpipe plot\_waterfall.pipe}\\
or (parallel executing by using $N$ MPI processes)

{\tt \$mpiexec -n N tlpipe plot\_waterfall.pipe}

For more details about the pipeline and input parameter file and
examples of more complex executing flows, readers can refer to {\tt
  tlpipe}'s document\footnote{\url{http://tianlaiproject.github.io/tlpipe}}.

\lstset{language=Python}
\lstset{frame=lines}
\lstset{caption={An example input parameter file for the pipeline}}
\lstset{label={lst:code_direct1}}
\lstset{basicstyle=\footnotesize}
\begin{lstlisting}
# -*- mode: python; -*-

# plot_waterfall.pipe


pipe_tasks = []
pipe_outdir = './output_waterfall/' # output path


import glob
data_dir = 'dir/to/data' # input data directory
files = sorted(glob.glob(data_dir+'/*.hdf5'))


# data selection
from tlpipe.timestream import dispatch
pipe_tasks.append(dispatch.Dispatch)
### parameters for Dispatch
dp_input_files = files # data files as list
dp_start = 0 # select data from the start
dp_stop = 450 # 4 seconds * 450 = 30 minutes
dp_out = 'dp'

# find and mask noise source signal
from tlpipe.timestream import detect_ns
pipe_tasks.append(detect_ns.Detect)
### parameters for Detect
dt_in = dp_out
dt_out = 'dt'

# plot waterfall of selected data
from tlpipe.plot import plot_waterfall
pipe_tasks.append(plot_waterfall.Plot)
### parameters for Plot
pwf_in = dt_out
pwf_flag_ns = True # mask noise source signal
pwf_fig_name = 'waterfall/wf' # figure name
pwf_out = 'pwf'
\end{lstlisting}

As a second example, the following input parameter file shows a
simplified map-making pipeline, where we have only included several
main task modules, and ingored tasks like data check, abnormal value
flagging, visualization and other things.

\lstset{language=Python}
\lstset{frame=lines}
\lstset{caption={Simplified input parameter file for map-making}}
\lstset{label={lst:code_direct2}}
\lstset{basicstyle=\footnotesize}
\begin{lstlisting}
# -*- mode: python; -*-

# map_making.pipe

pipe_tasks = []
pipe_outdir = './output_map/' # output path


import glob
data_dir = 'dir/to/data' # input data directory
files = sorted(glob.glob(data_dir+'/*.hdf5'))


from tlpipe.timestream import dispatch
pipe_tasks.append(dispatch.Dispatch)
### parameters for Dispatch
dp_input_files = files
dp_iterable = True
dp_iter_num = 5 # iterate for 5 days
dp_tag_input_iter = False
dp_out = 'dp'

from tlpipe.timestream import detect_ns
pipe_tasks.append(detect_ns.Detect)
### parameters for Detect
dt_in = dp_out
dt_iterable = True
dt_out = 'dt'

from tlpipe.timestream import rfi_flagging
pipe_tasks.append(rfi_flagging.Flag)
### parameters for Flag
rf_in = dt_out
rf_iterable = True
rf_first_threshold = 6.0
rf_tk_size = 1.0
rf_fk_size = 3.0
rf_out = 'rf'

from tlpipe.timestream import ns_cal
pipe_tasks.append(ns_cal.NsCal)
### parameters for NsCal
nc_in = rf_out
nc_iterable = True
nc_save_gain = True
nc_out = 'nc'

from tlpipe.timestream import rt2ts
pipe_tasks.append(rt2ts.Rt2ts)
### parameters for Rt2ts
r2t_in = nc_out
r2t_iterable = True
r2t_out = 'r2t'

from tlpipe.timestream import ps_cal
pipe_tasks.append(ps_cal.PsCal)
### parameters for PsCal
pc_in = r2t_out
pc_iterable = True
pc_calibrator = 'cyg' # calibrator
pc_span = 1200
pc_subtract_src = True
pc_apply_gain = True
pc_save_gain = True
pc_gain_file = 'cyg_gain.hdf5'
pc_temperature_convert = True
pc_out = 'pc'

from tlpipe.timestream import re_order
pipe_tasks.append(re_order.ReOrder)
### parameters for ReOrder
ro_in = pc_out
ro_iterable = True
ro_out = 'ro'

from tlpipe.timestream import accumulate
pipe_tasks.append(accumulate.Accum)
### parameters for Accum
ac_in = ro_out
ac_iterable = True
ac_out = 'ac'

from tlpipe.timestream import barrier
pipe_tasks.append(barrier.Barrier)
### parameters for Barrier

from tlpipe.timestream import average
pipe_tasks.append(average.Average)
### parameters for Average
av_in = ac_out
av_keep_last_in = True
av_out = 'av'

from tlpipe.timestream import map_making
pipe_tasks.append(map_making.MapMaking)
### parameters for MapMaking
mm_in = av_out
mm_auto_correlations = False
mm_pol = 'I' # make Stokes I image
mm_dirty_map = False
mm_method = 'tk' # Tikhonov regularization
mm_epsilon = 0.0001
mm_out = 'mm'
\end{lstlisting}

Finally, the header of a Tianlai HDF5 
data file is shown in Table \ref{tab:header} as an example. The meaning of each parameter 
is omitted here, though it is recognizable from the abbreviate variable names. Interested reader may consult the document on the code repository site 
and the report on Tianlai 
system \citep{2020SCPMA..6329862L} for further details.

\begin{table}[htbp]
  \centering
  \caption{Header of an example Tianlai HDF5 data file.}
  \begin{threeparttable}
    \begin{tabular}{l}
    \hline
    File: 20180322181759\_20180322182754.hdf5\tnote{*}\\
    \hline
/.attrs[``comment"]: Note potential instrumental RFI.\\
/.attrs[``observer"]: XXX\tnote{**}\\
/.attrs[``history"]: Recorded from the correlator.\\
/.attrs[``keywordver"]: 0.0\\
/.attrs[``sitename"]: Hongliuxia Observatory\\
/.attrs[``sitelat"]: 44.15268333\\
/.attrs[``sitelon"]: 91.80686667\\
/.attrs[``siteelev"]: 1493.7\\
/.attrs[``timezone"]: UTC+08h\\
/.attrs[``epoch"]: 2000.0\\
/.attrs[``telescope"]: Tianlai-Cylinder-I\\
/.attrs[``nants"]: 3\\
/.attrs[``npols"]: 2\\
/.attrs[``nfeeds"]: 96\\
/.attrs[``cylen"]: 40.0\\
/.attrs[``cywid"]: 15.0\\
/.attrs[``recvver"]: 0.0\\
/.attrs[``lofreq"]: 935.0\\
/.attrs[``corrver"]: 0.0\\
/.attrs[``samplingbits"]: 8\\
/.attrs[``corrmode"]: 1\\
/.attrs[``inttime"]: 3.99507456\\
/.attrs[``obstime"]: 2018/03/22 18:17:59.457007\\
/.attrs[``sec1970"]: 1521713879.46\\
/.attrs[``nfreq"]: 1008\\
/.attrs[``freqstart"]: 685.9765625\\
/.attrs[``freqstep"]: 0.1220703125\\
antpointing   shape =  (1, 96, 4)\\
antpointing.attrs[``unit"]: degree\\
blorder   shape =  (18528, 2)\\
channo   shape =  (96, 2)\\
feedno   shape =  (96,)\\
feedpos   shape =  (96, 3)\\
feedpos.attrs[``unit"]: meter\\
noisesource   shape =  (1, 3)\\
noisesource.attrs[``unit"]: second\\
nspos   shape =  (1, 3)\\
nspos.attrs[``unit"]: meter\\
pointingtime   shape =  (1, 2)\\
pointingtime.attrs[``unit"]: second\\
polerr   shape =  (96, 2)\\
polerr.attrs[``unit"]: degree\\
vis   shape =  (150, 1008, 18528)\\
weather   shape =  (1, 10)\\
    \hline
    \end{tabular}
    
    \begin{tablenotes}
      \footnotesize
      \item[*] Data is taken from 2018/03/22 18:17:59 to 2018/03/22/ 18:27:54 in Beijing Standard Time.
      \item[**] The name of the actual observer is not shown here.
    \end{tablenotes}
  \end{threeparttable}
  \label{tab:header}
\end{table}

\bibliographystyle{model2-names}
\bibliography{refs}
\end{document}